# A Learning Algorithm for Change Impact Prediction

Vincenzo Musco, Antonin Carette, Martin Monperrus, Philippe Preux
University of Lille and INRIA
Lille, France
Contact: vincenzo.musco@inria.fr

## ABSTRACT

Change impact analysis (CIA) consists in predicting the impact of a code change in a software application. In this paper, the artifacts that are considered for CIA are methods of object-oriented software; the change under study is a change in the code of the method, the impact is the test methods that fail because of the change that has been performed. We propose LCIP, a learning algorithm that learns from past impacts to predict future impacts. To evaluate LCIP, we consider Java software applications that are strongly tested. We simulate 6000 changes and their actual impact through code mutations, as done in mutation testing. We find that LCIP can predict the impact with a precision of 74%, a recall of 85%, corresponding to a $F$-score of 64%. This shows that taking a learning perspective on change impact analysis let us achieve good precision and recall in change impact analysis.

## 1. INTRODUCTION

Change impact analysis consists in predicting the impact of a code change in a software application [1, 2]. There are different kinds of impacts: the software modules that may be broken [1], the documentation that must be impacted [3], the developers that must keep informed about this change [4]. In this paper, we focus on the first kind: impacts at the code level.

There are two important challenges for change impact analysis. First, it must scale to the size of today's software systems, that often consists of thousands of modules. Second, the predictions must be accurate. The accuracy of impact prediction has two dimensions: whether predicted impacted elements are actually impacted (false positives) and whether actually impacted elements are all predicted (false negatives). This boils down to the classical information retrieval measures: precision, recall, and $F$-score. *Our motivation is to design a change impact prediction system that addresses both challenges: scaling to large systems while retaining a high accuracy (high precision, high recall, high $F$-score).*

We favor a learning perspective for the problem of change impact analysis. By learning, we mean that the change impact analysis could exploit previous actual impacts to form some kind of knowledge. This knowledge then guides the prediction of the future impacts. To this extent, a learning-based change impact analysis technique is the opposite of *a priori* approaches which design upfront the whens and hows the impact propagates, such as those based on program dependence graphs [5] to only name a few.

In this paper, we consider change impact analysis formulated as follows. The artifacts that are considered are methods of object-oriented software; the change under study is a change in the code of the method, the impact is the test methods that fail because of the change that has been performed. We propose an algorithm, called LCIP that learns from past impacts to predict future impacts. For instance, the algorithm may learn that a change in method `foo` is always followed by a failure observation in `bar`. Our algorithm is based on the call graph, where each node is a method and each edge a potential call between two methods, as observed statically. We consider a standard Class-Hierarchy-Analysis (CHA) call graph. The learning strategy consists of decorating each call graph edge with a weight between zero and one. This weight represents the likelihood of an edge to propagate an impact. The weight is updated based on actual impacts that are given as training data to the system.

To evaluate our system, we consider 2 Java software applications totaling 120,000+ lines of code. We simulate 6000 changes and their actual impact through code mutations, as done in mutation testing. We use ten-fold cross validation to measure the precision, recall and $F$-score of change impact prediction. We compare our system to two algorithms: one being a standard transitive closure on the call graph, the other one is a basic learning strategy that we introduce for the sake of understanding the actual learning that happens in LCIP. We find that LCIP can predict the impact with a precision of 74%, a recall of 85%, corresponding to a $F$-score of 64%. This validates our intuition that learning can be done for change impact analysis. In addition, our approach does not trade performance for accuracy: learning lasts in average 26 seconds and prediction 256 milliseconds. This indicates that our approach may scale-up to really-large systems.

To sum up, our contributions are:

- An algorithm for learning the impact of software changes. It is based on decorating call graph edges with a weight representing a likelihood to propagate a change. Mu-tants are used to learn those weights.

- An experimentation made on 2 popular Java programs, totaling 120,000+ lines of code. The experiment shows that a learning approach is a promising approach for change impact prediction.

- We publish all our code as open-source. The frameworks for mutation, impact prediction and learning can foster future research in this direction.

The remainder of this paper is structured as follows. In Section 2, we present concepts used in this paper. In Section 3, we introduce our learning approach. In Section 4, we present our framework and dataset. We also pose our research question and answer it by presenting and analyzing our experimental results. In Section 5, we present works related to our paper in fields such as machine learning, impact analysis and mutation testing. In Section 6, we conclude this paper.

## 2. BACKGROUND

### 2.1 Change Impact Analysis

Software is made of interconnected pieces of code (*e.g.* methods, classes, *etc.*). Through those connections, the effects of a change to a given part of the code can propagate to many other parts of the software. Those other parts, which can be potentially anywhere in the software, can then be impacted by the initial change, acting like a ripple. Change Impact Analysis (*a.k.a.* CIA) is defined as "the determination of potential effects to a subject system resulting from a proposed software change" [6].

In this paper, we use Bohner's definition of the basic software change impact analysis process [6]. Assuming a change has been performed, Bohner defines the following sets used in impact analysis: (i) the "starting impact set" ($SIS$) is the list of software parts which can be impacted by the change; (ii) the "candidate impact set" ($CIS$) (also called the "estimated impact set" [7]) is the list of software parts predicted as impacted by a change impact analysis technique; (iii) the "actual impacted set" ($AIS$) is the list of parts of the software which are actually impacted by the change; (iv) the "false negative impact set" ($FNIS$) is the list of missed impacts by the technique [1]; (v) the "false positive impact set" ($FPIS$) is the list of over-estimated impacts returned by the technique (*i.e.* false positives).

### 2.2 Mutation Analysis

Mutation analysis [8] consists of assessing test suite quality by applying minor changes to a software resulting in slightly modified versions of the software which are called *software mutants*. These changes result from the application of a *mutation operator* which describes what should be changed in the code, and how it should be modified.

A standard mutation testing scenario is that (i) the software tests are run on software mutants to ensure the change is noticed by at least one test and (ii) if no test kills a mutant, complementary tests are written.

---

[1]Bohner named this set the "discovered impact set" ($DIS$), but this naming is not appropriate in our context and may be confusing.

### 2.3 Call Graph

Grove et al. [9] define "*the program call graph [as] a directed graph that represents the calling relationships between the program's procedures (...) each procedure is represented by a single node in the graph. Each node has an indexed set of call sites, and each call site is the source of zero or more edges to other nodes, representing possible callees of that site*". This is the definition taken in this paper.

As we want to study Java applications which are object-oriented, we also take into consideration the class hierarchy analysis (*a.k.a.* CHA), *i.e.* the inheritance and usage of interfaces.

## 3. CONTRIBUTION

In this section, we describe the Learning Change Impact Prediction (LCIP) algorithm: a new learning algorithm for change impact analysis. This algorithm made of two phases is based on decorating call graphs with weights on the edges.

### 3.1 Approach Overview

We use the call graph to estimate the change and error propagation. Our key insight is that error propagation through a call graph edge is not systematic: some edges may propagate errors, others not.

Thus, we propose a stochastic call graph by adding weights on the call graph edges. Those weights are first learned from existing data, and then used to compute impacts. These weights range from 0 to 1 where 0 means that the edge never propagates the error, while 1 means that the edge always propagates it. Any value in between means the propagation sometimes occurs and sometimes not. The initial weights are set to 0 meaning that we start by assuming that no impact is propagated at all. In this paper, we consider that the impacted nodes are always test methods; this is motivated by the fact that broken assertions unambiguously denote impacted behavior.

Our approach is thus composed of two distinct phases. The **learning phase** consists of learning the weights based on a set $M$ of changes and their impacts. In our graph-based context, a change is modeled as a modified node and an impact is a set of nodes whose behavior is impacted by the change. The **prediction phase** is when, given a call graph and a specific change in a method by a developer (*i.e.* on a node of the call graph), an impact prediction algorithm computes the candidate impacted set $CIS$ composed of all test nodes using the weights. The prediction represents all tests that may be broken by the change.

Figure 1 is an example of a call graph. Three types of nodes are presented: application nodes (plain circle), test nodes (circle with a T) and the changed node (double circle). We have four methods: `mul` which computes a multiplication and two methods depending on it: `pow` and `fac` which computes respectively the power and the factorial. Then, a test function is defined for each method. The `op` method uses reflection mechanism to call `mul`, which is not resolved statically, resulting in the absence of edge between `op` and `mul`.

Consider that an error is introduced by a developer in the `mul` method. Bohner's sets are illustrated in the figure. The $SIS$ set is made of all software test nodes. The top part of the horizontal line shows the $SIS$ set. The $AIS$ set (thin rectangle) is composed of the test nodes which fail when actually running the software tests. The $CIS$ set (thick rectangle) is composed of the candidate test nodes determined by a virtual change impact prediction technique (in this example it is an arbitrary set). The $FNIS$ set (dashed rectangle) shows a false negative test due to the absence of edge in the call graph. The $FPIS$ set (dotted rectangle) shows a false positive case: the `pow` method seems to be impacted according to the change impact analysis technique (as it is part of the $CIS$ set), but when running the tests, it did not fail.

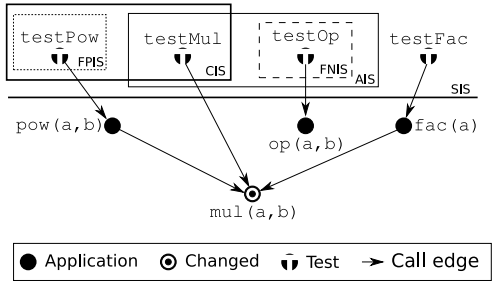

**Figure 1: Example of a call graph in which a change has been introduced. The class graph includes application nodes (bottom), test nodes (top) and call graph edges. The rectangles illustrate Bohner's sets used in Change Impact Analysis.**

### 3.2 Learning

The LCIP learning phase requires as input data: (i) a call graph $G$ such as described in Section 2.3; (ii) a set $M$ made of pairs containing changed methods and the impacted nodes (other methods) such as $M = \{\{m, AIS\} \mid \forall S\}$, where $S$ is different versions of the software composed with the method $m$ on which the change has occurred and the set of methods $AIS$ which have been impacted by the change on run time.

$M$ can be seen as a set of "examples" that our impact learning algorithm uses as its input. The algorithm estimates the weights to assign to the edges of the call graph in order to predict accurately the propagation of changes. Our call graph weight learning algorithm is shown on Algorithm 1.

---

**Algorithm 1:** The call graph weight learning algorithm using call graph. *update_weight* is a sub-algorithm which updates the weights.

**Input**: $G$ the call graph and $M$ the data composed of changed points and their actual impacts.
**Output**: a weighted graph
1 **begin**
2   $L \leftarrow G$ with $weights = 0$ for all edges
3   **for** *each* $\{m, AIS\} \in M$ **do**
4     **for** *each* $t \in AIS$ **do**
5       $update\_weight(m, t, L)$
6   **return** $L$;

---

For all changed node $m$ in the program (line 3) and each actually impacted node $t$ (line 4), we update the weight of edges belonging to all paths from $m$ to $t$ following an update algorithm (line 5). In this paper, we propose two algorithms for updating the weights.

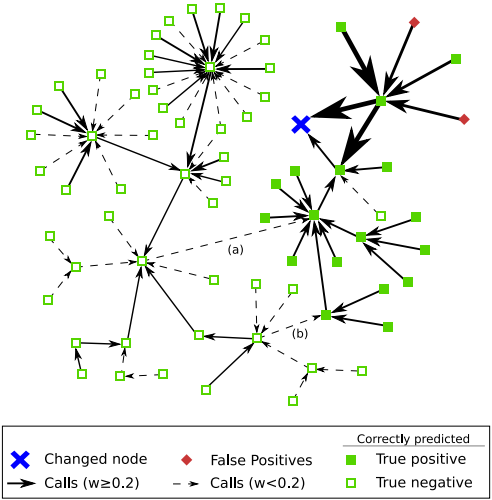

**Figure 2: An illustration of impact prediction based on weighted call graphs. Edges with a low weight ($< 0.2$) are considered as non-propagating the impact of the change.**

Figure 2 illustrates our impact analysis technique based on decorating call graphs with weights. It is based on real data obtained in our experiments. Assume that a change has occurred at the method denoted by a blue cross. The edges are method calls, and the thickness of the edges represents the weight of the edge after learning. A thicker edge means a weight close to 1 and a thinner means a weight close to 0.2. The dashed edges are those which have a weight smaller than 0.2 and which are not considered for propagation. The green squares and red diamonds represent nodes predicted as impacted by our approach. The prediction is of high accuracy because the weights of the two edges (a) and (b) is low after learning. Consequently, the impact of the change is stopped and does not ripple to the left-hand side of the graph. On the contrary, a basic impact prediction based on a transitive closure on the call graph would predict far too much tests (all the true negatives would become false positives).

#### 3.2.1 Binary Update Algorithm

This algorithm assumes a binary impact propagation: impact is propagated or not. Thus, the model consists in assigning 0 or 1 to the edge weights as follows. If at least one impact has been observed between a graph node and the changed node, then all edges belonging to all paths going from the former to the latter are labeled as 1, otherwise, these edges are labeled with a 0. Thus, this approach considers that if an edge has once propagated an error, it will always do so (as 1 is always greater or equal to the threshold, no matter its value). Algorithm 2 formalizes this idea. This algorithm is deterministic. Thus, running the same algorithm on the same data several times will always produce the same results.

#### 3.2.2 Dichotomic Update Algorithm

We now explore a more realistic model where error propagation is conditioned by the current state (*e.g.* the error propagates if coming from the if-branch of a condition, and

**Algorithm 2:** Algorithm Binary for updating the edge weights in paths between a node and the changed point.

**Input**: $m$ the node which has changed, $t$ the considered impacted node and $L$ the weighted call graph
**Output**: the weights of $L$ are updated
1 **begin**
2    **for** *each edge in all paths from $m$ to $t$ in $L$* **do**
3       $w_{edge} \leftarrow 1$

not propagated if coming from the else branch). This means that some edges propagate errors but only sometimes, in particular cases. This is represented by a weight that is neither 0 nor 1 but in between.

The Dichotomic algorithm updates the weights according to an estimation of the probability that a node would be broken by a change. This estimation is based on training data as follows: $p_{t,m} = \frac{\alpha_t}{\beta_m}$, where $\alpha_t$ is the number of times the node $t$ is impacted over all changes occurring on the same method $m$ and $\beta_m$ is the number of times the method has been changed.

The idea of Algorithm Dichotomic (Algorithm 3) is to slowly converge to $p_{t,m}$ example after example. For each training example, the weight $w$ of each edge which belongs to a path between a changed node $m$ and an impacted node $t$ is computed in a dichotomic way: the new weight is the mean value between the current weight and the empirical probability.

Note that this approach cannot be used in line as we require a set of changes and their impact to compute future weights.

**Algorithm 3:** Algorithm Dichotomic for updating the edge weights between a node and the changed point.

**Input**: $m$ the node which has changed, $t$ the considered impacted node, $p_{t,m}$ the empirical probability for $(m,t)$ and $L$ the weighted call graph
**Output**: the weights of $L$ are updated
1 **begin**
2    **for** *each edge in all paths from $m$ to $t$ in $L$* **do**
3       $w_{edge} \leftarrow (w_{edge} + p_{t,m})/2$

## 3.3 Prediction

At prediction time, LCIP is based on Algorithm 4. This algorithm takes as input the node $n$ corresponding to the point in the code where the change has occurred and a threshold value $th$ lying in the range $[0,1]$. It returns a candidate impacted set $CIS$ composed of all nodes predicted as impacted. To do so, starting at the node being changed, the graph edges are followed to determine which nodes can be reached. The weights are used to prune some edges which are unlikely to propagate the change, according to the threshold value (line 6). If the weight is lower than the threshold value, the error does not propagate across the edge; otherwise the edge propagates the error.

## 4. EVALUATION

We now present the evaluation of LCIP for impact prediction. We are especially interested in answering the following research question: *To what extent LCIP improves the accuracy of impact prediction?*

We want to determine whether our prediction algorithm based on learning can improve the prediction scores compared to the use of a baseline transitive closure prediction technique described in 4.1.2.

**Algorithm 4:** The impact prediction algorithm that uses the learned weights of the call graph

**Input**: $L$ the weighted call graph, $n$ the changed node, $th$ the threshold
**Output**: the set of nodes which are predicted as impacted
1 **begin**
2    $CIS \leftarrow \{\}$
3    **for** *each node $i$ connected to $n$ in $L$* **do**
4       **if** *$i$ is not visited* **then**
5          mark node $i$ as visited
6          **if** $w_i >= th$ **then**
7             $CIS \leftarrow CIS \cup \{i\}$
8             $CIS \leftarrow CIS \cup visit(i)$
9    **return** $CIS$

## 4.1 Experimental Protocol

We explain how we evaluate our approach, the dataset and the configuration parameters we use. In this paper, we use mutation injection to simulate changes in the software in order to learn the error propagation profile.

The evaluation follows several steps: (i) we create mutants for a software application, they simulate changes; (ii) we extract the corresponding call graphs with and without class hierarchy analysis; (iii) we split the dataset of mutants in a training set and a testing set; (iv) we run our learning algorithm based on the mutants of the training set. This results in a weighted call graph.

Then, for each mutant of the testing set: (i) we compute the actual impact set by running the original test cases that come with the software package under study; (ii) we predict the impact set for each mutant with our technique, using the weights learned in the previous step; (iii) we compute performance metrics by comparing the predicted impact set and the actual impact set.

The idea of creating mutants is to create synthetic changes that have an impact observable in test cases, this is further discussed in 4.3. In addition, we use 10-fold cross-validation [10]: for each software, we partition the mutants into 10 subsets of equal size. We take 9 subsets to train the model (with Algorithm 1) and the one remaining is used to assess the model (with Algorithm 4). This process is run 10 times. We compute the mean value of the evaluation metrics considered over these 10 runs. For Dichotomic, we use a threshold value of 0.2 for prediction. This value is the best one according to a systematic grid search of all thresholds ranging from 0 to 1 with an increment of 0.1.

Our mutation, learning and evaluation framework is open-source and freely available online[2].

---
[2]https://github.com/v-m/PropagationAnalysis, https://github.com/k0pernicus/PropL

Table 1: Precision, recall and $F$-score obtained for change impact prediction with three techniques: transitive closure (TC), Algorithm Binary and Algorithm Dichotomic. The bold faced values indicate the best results for a single metric (up to rounding precision).

| Package | Op. | #mut. | Precision | | | Recall | | | F-score | | |
|---|---|---|---|---|---|---|---|---|---|---|---|
| | | | TC | Bin | Dic | TC | Bin | Dic | TC | Bin | Dic |
| Collections | ABS | 600 | 0.55 | 0.75 | **0.81** | 0.80 | 0.80 | **0.83** | 0.43 | 0.59 | **0.67** |
| | AOR | 600 | 0.65 | 0.78 | **0.81** | 0.68 | **0.70** | 0.63 | 0.44 | **0.55** | 0.51 |
| | LCR | 600 | 0.56 | 0.69 | **0.72** | 0.77 | **0.78** | 0.77 | 0.43 | 0.51 | **0.53** |
| | ROR | 600 | 0.59 | 0.80 | **0.84** | 0.64 | **0.68** | 0.66 | 0.40 | 0.55 | **0.56** |
| | UOI | 600 | 0.63 | 0.80 | **0.81** | 0.68 | **0.70** | **0.70** | 0.44 | **0.57** | 0.56 |
| | All | 3000 | 0.60 | 0.76 | **0.80** | 0.71 | **0.73** | 0.72 | 0.43 | 0.55 | **0.57** |
| Lang | ABS | 600 | 0.30 | 0.68 | **0.69** | **0.99** | **0.99** | 0.98 | 0.33 | 0.69 | **0.70** |
| | AOR | 600 | 0.47 | 0.65 | **0.67** | **0.99** | **0.99** | **0.99** | 0.52 | 0.68 | **0.70** |
| | LCR | 600 | 0.38 | 0.55 | **0.59** | **0.98** | 0.96 | 0.93 | 0.43 | 0.59 | **0.62** |
| | ROR | 600 | 0.49 | 0.66 | **0.67** | **0.98** | 0.97 | 0.97 | 0.54 | 0.68 | **0.69** |
| | UOI | 600 | 0.51 | 0.74 | **0.75** | **0.98** | **0.98** | 0.97 | 0.55 | 0.75 | **0.77** |
| | All | 3000 | 0.43 | 0.66 | **0.67** | **0.98** | **0.98** | 0.97 | 0.47 | 0.68 | **0.70** |
| Total | | 6000 | 0.52 | 0.71 | **0.74** | 0.85 | **0.86** | 0.85 | 0.45 | 0.62 | **0.64** |

### 4.1.1 Evaluation Metrics

To assess and compare the performance of our impact prediction techniques, we use metrics computed on $CIS$ and $AIS$ presented by Arnold and Bohner [7].

The size of the $CIS$ has to be as close as possible to the size of the $AIS$ as it quantifies the amount of elements retrieved by the impact analysis. We express this in terms of precision, recall and $F$-score [11]. There is one precision, recall and $F$-score per mutant of the testing set. We then compute average precision, recall and $F$-score over all mutants of a fold. The *precision* $P$ is the proportion of predicted tests which are actually impacted;. The *recall* $R$ is the proportion of truly impacted tests that are retrieved. $P$ and $R$ are computed using Equation (1).

$$P = \frac{|AIS \cap CIS|}{|CIS|}; R = \frac{|AIS \cap CIS|}{|AIS|}. \qquad (1)$$

To make comparison easier, the $F$-score $F$ is the harmonic mean of $P$ and $R$. Since precision, recall and $F$-score are computed for each mutation, it is necessary to consider a mean value over all mutants of the testing set. *Our key goal is to improve the $F$-score of impact prediction as it takes into consideration both precision and recall.*

### 4.1.2 Baseline

As a baseline for change impact analysis, we compute the transitive closure of call graph nodes from the mutated node [1]. The result of the transitive closure is a list of all nodes potentially impacted by the change. Since the impact is only computed on test nodes, we remove application nodes from the impacted nodes in the transitive closure.

### 4.1.3 Dataset and Setup

In this paper, we have selected 2 *Java* software packages from the Apache Commons family: *Apache Commons Lang 3.5* (git #6965455) and *Apache Commons Collections 4.1* (svn r1610049). Using the open-source tool `cloc`, we computed a total of 122590 LOC for both projects (67509 for Lang and 55081 for Collections). The first one contains 2657 tests and the second one 5262 tests. The call graph size is 6195 nodes for Lang and 6271 for Collections with 9653 edges for Lang and 12130 for Collections.

### 4.1.4 Mutation Operators

We consider the 5 classical mutation operators validated by King and Offutt [12] adapted to the Java language (as those operators are firstly intended for Fortran). As shown by Offutt et al., these five mutation operators are sufficient to effectively implement mutation testing [13]. The five mutations operators are: (i) *Absolute value insertion (ABS)* in which each numerical expression or literal is replaced by its absolute value. (ii) *Arithmetic operator replacement (AOR)* in which each arithmetic expression is replaced by a new one with a different operator but same operands. (iii) *Logical connector replacement (LCR)* in which each logical expression is replaced by a new one with a different operator but same operands. (iv) *Relational operator replacement (ROR)* in which each relational expression is replaced by a new one with a different operator but same operands. (v) *Unary operator inversion (UOI)* where each arithmetic expression `x` is mutated to its opposite value (*i.e.* `x * -1`), its incremented value (*i.e.* `x + 1`) and its decremented value (*i.e.* `x - 1`). Moreover, each logical expression `x` is mutated to its complemented form (*i.e.* `!x`). For AOR and LCR, the whole expression can also be mutated to `true` or `false` constant. For LCR and ROR, possible mutations are the left or right operand alone.

## 4.2 Results

In this section, we answer our research question: *To what extent LCIP improves the accuracy of impact prediction?* All the raw empirical data is publicly available online[3].

Table 1 gives the values of the evaluation metrics presented in Section 4.1.1. The first, second, and third columns are respectively the name of the package, the mutation operator and the number of mutants considered. Then we have three multi-columns, one for each metric (precision, recall and $F$-score). Each multi-column is made of three columns which are the value obtained using the transitive closure (TC) baseline, the value obtained with the Binary algorithm and the value obtained with the Dichotomic algorithm. These values are the average over ten-fold cross validation. For each multi-column, the best value is shown in bold face.

---
[3] https://github.com/v-m/PropagationAnalysis-dataset

First, if we compare the performance of the transitive closure algorithm (TC) and the two learning algorithms, we observe that in 100% of cases, both the precision and the $F$-score are improved using our learning algorithms. For the recall, regarding Collection, our learning algorithms give better recall scores. For Lang, our learning algorithms give similar scores than TC with respect to 3/5 mutation operators. This means that in 8 cases out of 10, our learning algorithms give better recall scores than TC.

For the precision, the best improvement is for the Lang project with the ABS operator where the precision raises from 0.30 for TC to 0.68 using Binary algorithm and to 0.69 using the Dichotomic algorithm. Overall, the Binary and the Dichotomic algorithms respectively provide an average precision improvement of 0.20 and 0.22 when averaged over all projects and all mutation operators.

For the recall, the values are quite stable, the best improvement for Binary and Dichotomic algorithms are both with Collections. For Binary, the best improvement is obtained with the ROR operator for which the recall is improved from 0.64 for the baseline to 0.68 (+0.04). For Dichotomic, the best improvement is obtained with the ABS operator for which the recall is improved from 0.80 for the baseline to 0.83 (+0.03). However, there is no real improvement with our learning algorithm compared to the transitive closure baseline on average ($\pm 0.01$).

For the $F$-score, the best improvement is also obtained on the Lang project with the ABS operator where the $F$-score raises from 0.33 using the transitive closure baseline to 0.69 (+0.36) using the Binary algorithm, and 0.70 (+0.37) using the Dichotomic approach. Globally, the Binary and the Dichotomic algorithms lead respectively an average $F$-score improvement of 0.17 and 0.18. Figure 3 shows a scatter plot of the average $F$-scores obtained for all projects and mutation operators for TC ($x$-axis) and Dichotomic ($y$-axis). The line represents $y = x$. Since all points are in the upper-left part, above $y = x$, this figure graphically highlights that Dichotomic much improves the prediction. In addition, we also graphically note that some operators are really far from $y = x$, which means a strong improvement.

If we compare the two learning variants (Dichotomic and Binary), we see that the Dichotomic algorithm has better precision values than the Binary for 100% of the cases. But, Dichotomic only has better recall in 1 case out of 10 (in 3 cases they have the same score). $F$-scores are also better for the Dichotomic algorithm in 8 cases out of 10. The fact that Binary gives better recall while Dichotomic gives better precision is linked to the fact that Binary has a "all or nothing" spirit. If an edge is activated by only one mutant at learning time, then it will be considered as important as an other edge activated several times, by several mutants. To the opposite, with Dichotomic, rarely visited edges have a low weight, and the threshold prunes them at prediction time. Consequently, Binary considers more edges compared to Dichotomic, yielding more predicted nodes, hence higher recall and lower precision.

To assess the significance of the performance difference between Dichotomic and Binary, we run a Mann-Whitney-Wilcoxon statistical test on precision, recall and $F$-score over all mutants. The null hypothesis ($H_0$) is: "the metrics (either precision, or recall, or $F$-score) obtained for the Binary and the Dichotomic algorithms are drawn from the same population". Using the usual significance level $\alpha = 0.05$, the p-values are: $10 \times 10^{-16}$ for precision, $4.1 \times 10^{-12}$ for recall, and $8.7 \times 10^{-14}$ for $F$-score. For each metric, the conclusion is that the null hypothesis may be rejected, which means that Dichotomic is better in a statistically significant sense.

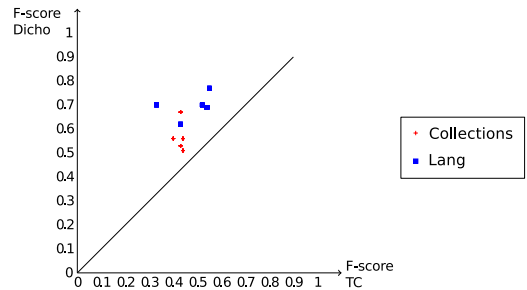

**Figure 3:** The performance improvement of the Dichotomic algorithm over the basic TC impact prediction. One point represents a mutation operator for a given project and is located at coordinates $(x = F_{TC}, y = F_{Dichotomic})$.

> To sum up, LCIP achieves better results for change impact prediction compared to a standard transitive closure approach. Among the two algorithms we propose in this paper, the best algorithm is the Dichotomic approach.

### 4.3 Threats to Validity

**Synthetic Changes** In our evaluation, we use synthetic changes for exploring the performance of our impact prediction algorithm. Our motivation for using synthetic changes is to have a large amount of data, which is necessary for learning. Another option would be to use impact of real code changes, however, those are extremely difficult to obtain, because developers never commit a change that breaks a test case. This is the reason why the related work only considers a very small evaluation benchmark. For instance, the evaluation of SENSA [14] uses 27 changes, which is much smaller than the 6000 simulated changes of our experiments. Also our results are dependent on the choice of synthetic mutation operators.

**Internal Validity** Our results are of computational nature. A major bug in our software can invalidate our findings. We have published all our code on Github so as to facilitate reproduction and falsification of our results, if necessary.

**External Validity** In this paper, we use 600 mutants per mutation operator and per project. Using ten-fold cross validation, this results in testing sets of up to 60 items. Since the aggregate performance measures (precision, recall and $F$-score) are rather stable over folds, we have confidence that this is enough to back up our conclusions. However future work with more mutants is required.

The impact prediction much depends on the structure of the call graphs [15]. For instance, the presence of large utility methods, with many incoming and outgoing edges has a direct impact on the prediction performance. Our results may only be valid for Java software, or even worse only valid for the projects under study. Future work in this field will strengthen the external validity of our findings.

## 5. RELATED WORK

Strug and Strug [16] use data and control flow graphs and classification for detecting similar mutants. They use similar tools as us (learning and mutation) to reduce the number of mutant considered when doing mutation testing. In our work, we use learning for change impact analysis. Do and Rothermel [17] described a protocol to study test case prioritization techniques based on mutation. Their protocol and ours share the same intuition, use test cases to determine which tests is impacted by the change. However, we have a different goal, they study test case prioritization, we study impact prediction. Hattori et al. [11] have used an approach based on call graphs to study the propagation. Their evaluation is made on a dataset made of three projects. Their goal was to show that the precision and recall are good tools as evaluation of the performance for an impact analysis technique. In contrast, we propose a concrete impact analysis technique. Law and Rothermel [18] have proposed an approach for impact analysis; their technique is based on a code instrumentation to analyze execution stack traces. They compare their technique against simple call-graphs on one small software. Our technique is compared to similar graphs but with two different software packages. Gethers et al. [19] also address impact analysis. However, they consider a slightly different problem setting, because they take as input a bug report or a modification request and not a single source code element as we do.

## 6. CONCLUSION

In this paper we have explored the possibilities of using learning to improve impact prediction based on call graphs. We have shown that we can improve the prediction performance with a learning algorithm based on the addition of weights to the call graph edges. From a software engineering perspective, our results show that having a learning perspective on change impact prediction is a promising research direction. Our future work focuses on improving the external validity, in particular with learning on more projects and mutants.